\documentclass[conference]{IEEEtran}
\IEEEoverridecommandlockouts
\usepackage{cite}
\usepackage{amsmath,amssymb,amsfonts}
\usepackage{algorithmic}
\usepackage{graphicx}
\usepackage{textcomp}
\usepackage{xcolor}
\def\BibTeX{{\rm B\kern-.05em{\sc i\kern-.025em b}\kern-.08em
    T\kern-.1667em\lower.7ex\hbox{E}\kern-.125emX}}
\begin{document}

\title{Industrial Machines Health Prognosis using a Transformer-based Framework}

\makeatletter
\newcommand{\linebreakand}{%
  \end{@IEEEauthorhalign}
  \hfill\mbox{}\par
  \mbox{}\hfill\begin{@IEEEauthorhalign}
}
\makeatother

\author{
\IEEEauthorblockN{1\textsuperscript{st} David J Poland}
\IEEEauthorblockA{\textit{Department of Computer Science} \\
\textit{University of Hertfordshire}\\
Hatfield, UK \\
d.j.poland@herts.ac.uk}

\and

\IEEEauthorblockN{2\textsuperscript{nd} Lemuel Puglisi}
\IEEEauthorblockA{\textit{Department of Computer Science} \\
\textit{University of Catania}\\
Catania, IT \\
lemuel.puglisi@phd.unict.it}

\and

\IEEEauthorblockN{3\textsuperscript{th} Daniele Ravì}
\IEEEauthorblockA{\textit{Department of Computer Science} \\
\textit{University of Hertfordshire}\\
Hatfield, UK \\
d.ravi@herts.ac.uk}
}
\maketitle 

\begin{abstract}
This article introduces Transformer Quantile Regression Neural Networks (TQRNNs), a novel data-driven solution for real-time machine failure prediction in manufacturing contexts. Our objective is to develop an advanced predictive maintenance model capable of accurately identifying machine system breakdowns. To do so, TQRNNs employ a two-step approach: (i) a modified quantile regression neural network to segment anomaly outliers while maintaining low time complexity, and (ii) a concatenated transformer network aimed at facilitating accurate classification even within a large timeframe of up to one hour. We have implemented our proposed pipeline in a real-world beverage manufacturing industry setting. Our findings demonstrate the model's effectiveness achieving an accuracy rate of 70.84\% with a 1-hour lead time for predicting machine breakdowns. Additionally, our analysis shows that using TQRNNs can increase high-quality production, improving product yield from 78.38\% to 89.62\%. We believe that predictive maintenance assumes a pivotal role in modern manufacturing, minimizing unplanned downtime, reducing repair costs, optimizing production efficiency, and ensuring operational stability. Its potential to generate substantial cost savings while enhancing sustainability and competitiveness underscores its importance in contemporary manufacturing practices.
\end{abstract}


\section{Introduction}
Predictive maintenance has emerged as a transformative approach with immense potential to revolutionize the prevention of machine failures in the manufacturing sector. It is commonly employed to anticipate and prevent unexpected machine failures and product damage. As industries progress towards Industry 4.0 \cite{Frank}, reducing manual labour and eliminating outdated maintenance practices have become essential for optimizing performance. Indeed, unplanned machine downtime can adversely affect performance and hinder the achievement of predefined goals. In fact, any manufacturing sector may encounter substantial challenges, including high production costs, quality issues, and reduced overall equipment effectiveness, leading to significant revenue losses (often in the millions of USD) if this problem is not accurately prevented \cite{Monia}.

Traditional maintenance strategies used today are based mainly on calendar-based assumptions or rely on statistical anomaly detection, which have many limitations. For instance, time-based total preventative maintenance involves routine component replacements, regardless of the actual system health. Such practices lead to costly interventions and missed production opportunities, without necessarily circumventing any unexpected failures. In contrast to traditional practices, predictive maintenance, as described by \cite{Li} and \cite{Lai}, proposes an adaptive alternative by integrating a large set of hardware sensors into the systems, which can identify anomalies through computational model-based methods.

In machinery maintenance and predictive analytics, data-driven modeling techniques have been developed to analyze real-time events within industrial equipment and, deep neural networks have played a pivotal role \cite{Serradilla}. Such approaches can process visual and raw signal data from operational machines, allowing for more accurate and timely predictions of maintenance needs and potential issues \cite{Yurek, Ma}.

Despite these advances, open challenges remain. Large datasets can introduce data corruption, affecting real-time monitoring accuracy \cite{Yang}. Harsh industrial settings and human interactions further intensify predictive maintenance complexities, and researchers have sought to amalgamate modeling techniques tailored to specific machines.

In this paper, we introduce TQRNNs, a novel transformer-based framework specifically developed and tested within the practical context of the beverage industry. Our methodology is inspired by and built upon the work proposed in \cite{Grigsby}. It incorporates quantile processing methods for extracting features from complex machinery signals, allowing the analysis of time-sequence data directly using transformer networks. We assessed the proposed solution through live trials across multiple real-industry production lines, and the results show that our model leads to improvements in machine prognostic health with respect to state-of-the-art solutions. This enables the pre-emptive identification of defective products, ensuring only high-quality items reach consumers. The contributions of our article are twofold: i) we introduce a novel methodology for predicting machine failures by leveraging edge sensor data, quantile regression neural network and a transformer-based framework and ii) we evaluate the effectiveness of our model in identifying malfunctioning machines in a real-world production environment.

While we have demonstrated the application of our pipeline specifically in the context of beverage cans production, we firmly believe that our approach holds immense potential for a wide range of applications across diverse sectors. This pipeline offers substantial opportunities to significantly enhance operational efficiency, minimize costly disruptions through optimized maintenance scheduling, prolong the lifespan of critical assets, and reduce machine downtime. By harnessing the power of our predictive maintenance solution, industries can unlock substantial gains in productivity, cost savings, and overall operational excellence.

The rest of the article is structured as follows: Section \ref{Related_Work} reviews related work. Section \ref{architecture} details the proposed pipeline. In Section \ref{experiment}, we discuss the experimental setup and present the evaluation results. Finally, Section \ref{conclusion} provides the conclusions.

\section{Related Work}\label{Related_Work}
Research in predictive maintenance within the manufacturing sector has focused on developing advanced systems and techniques to address production losses, machine outages, and equipment failures. The main goals are to minimize manufacturing output reduction, increase product quality, and optimize maintenance schedules. These efforts have leveraged various analytical models, machine learning techniques, deep learning architectures, and innovative data analysis approaches.

Early work explored analytical parametric predictive maintenance decision frameworks that leverage system health prognosis results to inform maintenance decisions. For instance, research by \cite{Huynh} proposed a parametric predictive maintenance decision framework that utilizes the precision of the Remaining Useful Life (RUL) estimate as a condition index to schedule maintenance interventions like inspections and replacements for stochastically deteriorating systems. Traditional machine learning techniques have also been employed for fault prediction and RUL estimation of assets. This includes the use of Support Vector Regression (SVR) \cite{Hanshuo}, Regression Tree \cite{Bakir}, and Decision Forest Regression \cite{Yurek}. However, these models often struggle to capture the intricate relationships and complex dynamics inherent in interconnected manufacturing systems, particularly when dealing with high-dimensional and time-series data. To address these limitations, researchers have explored the application of deep learning models, such as Recurrent Neural Networks (RNNs) and Long Short-Term Memory (LSTM) architectures \cite{Rivas, Xiao}. These neural network architectures have demonstrated improved long-term prediction accuracy by effectively handling time-series data collected from manufacturing processes. Furthermore, hybrid models combining deep autoencoders with bidirectional LSTMs have been proposed \cite{Chen_2021} to accurately estimate system health states and RUL, enabling dynamic predictive maintenance scheduling tailored to modern engineered systems with multiple sensors.

While RNNs and LSTMs have marked significant progress, they can encounter issues with long-term dependencies due to vanishing and exploding gradients, limiting their effectiveness in more complex or extended manufacturing scenarios. To overcome these challenges, recent research has focused on transformer models leveraging self-attention mechanisms. A notable example is the Spacetimeformer \cite{Grigsby}, a transformer-based model that treats multivariate forecasting as a "spatiotemporal sequence" problem, where each input token represents the value of a specific variable at a given time. This approach enables the transformer to simultaneously process and integrate interactions among spatial, temporal, and value dimensions, effectively managing long-range dependencies in complex datasets.

In addition to these techniques, researchers have also explored complementary strategies to enhance predictive maintenance capabilities. These include anomaly detection and unsupervised learning approaches \cite{Devlin}, real-time data collection from edge sensors \cite{Ma}, integration of meta-modeling data \cite{Batur}, rapid data analysis methods \cite{Korkmaz}, and hybrid approaches combining multiple predictive models \cite{Tan}. These efforts reflect the sector's movement towards more sophisticated, accurate, and efficient predictive maintenance techniques tailored to the unique demands of smart manufacturing systems. Despite these recent advancements, there remains a need for robust and accurate predictive maintenance approaches that can effectively handle multivariate time-series data with complex dependencies. For this reason, we proposed TQRNN, a novel pipeline aimed at addressing these challenges by providing a framework for predictive maintenance tailored to handle multivariate time-series data with intricate dependencies.

\section{System Architecture: Proposed TQRNN pipeline}\label{architecture}
In this section, we present our novel data-driven framework called TQRNNs. The proposed pipeline is depicted in Figure~\ref{fig:Pipeline}. It starts with an input block that processes data from 43 sensors, followed by a set of 10 $\times$ 43 Quantile Regression Neural Networks (QRNNs). A secondary layer of 2 $\times$ 43 QRNNs (2nd phase) is then used to further refine the preliminary results and generate two distinct outputs for each sensor. The process culminates with a final transformer network that identifies possible anomalies in the data. We will describe each of these blocks in the sections below.

\begin{figure*}[t]
\begin{center}
\includegraphics[width=0.8\textwidth,trim={0 12cm 9cm 0},clip]{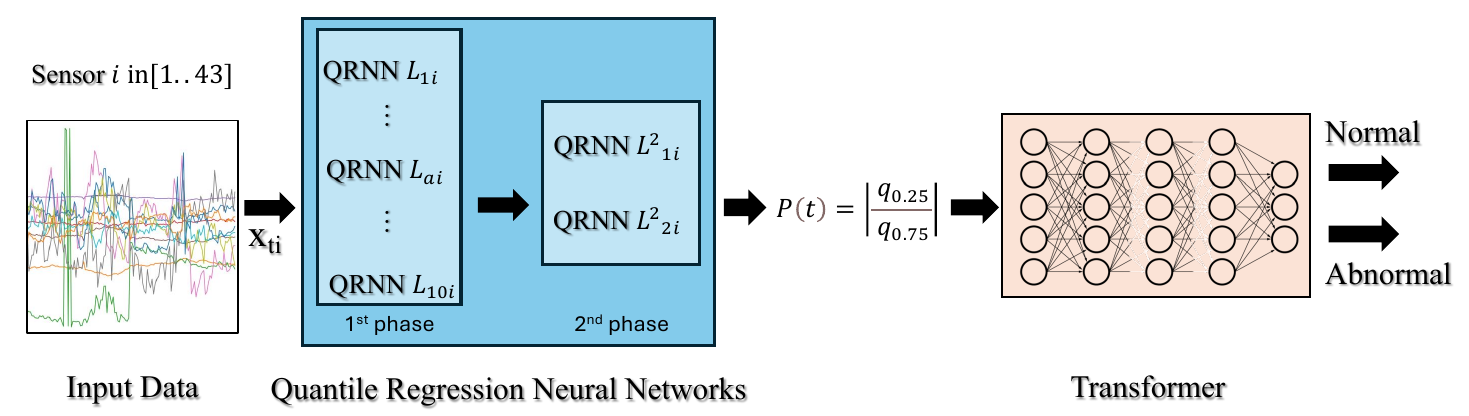}
\caption{Proposed Pipeline TQRNNs: Input data from 43 sensors is processed by a series of QRNNs, first with  10×43 networks, followed by 2×43 QRNNs. The final block employs a transformer network to synthesize outputs and predict machine anomalies.\label{fig:Pipeline}}
\end{center}
\end{figure*} 

\subsection{Dataset}
The dataset utilized in this study was collected from eight different machines within a beverage Manufacturing facility., involving a total of 43 unique sensors installed within each machine. The data were manually categorized into two distinct groups: "Normal" and "Abnormal." The "Normal" category encompasses data representing the machine's operation under standard conditions, without any defects or breakdowns, and includes approximately 100,000 data points per sensor. Conversely, the "Abnormal" category contains data indicative of machine failures or production issues, also comprising roughly 100,000 data points for each of the 43 sensors.

We organized the dataset into three subsets for our analysis: a training set, a validation set, and a test set. The training set consists of 60\% of both normal and abnormal data points for each sensor. The validation set and the test set each contain 20\% of the normal and abnormal data points per sensor. This division ensures a comprehensive evaluation of our models across various conditions and scenarios.

\subsection{Input data}
The set of sensors used in our study are 6 three-axis vibration sensors positioned deep within the machine, along with 37 monitoring sensors comprising 7 flow sensors, 18 pressure sensors, 8 sensor health indicators, and 4 temperature sensors.

\subsection{Quantile Regression Neural Networks}
Quantile Regression Neural Networks (QRNNs) \cite{Cannon} are a class of neural network models designed to provide a comprehensive understanding of data by predicting multiple quantiles of the target variable's distribution. These neural networks go beyond traditional point estimates as they excel at modeling complex non-linear relationships, and they are inherently robust to outliers, making them particularly well-suited for time-series data such as raw sensor data \cite{Biau}. Building upon the QRNN proposed in \cite{Mohammad}, the initial step in our pipeline involves employing an ensemble of these QRNNs, represented as $\mathcal{L}_{ai}$, operating on the data from each sensor $i = 1, \dots, 43$. In particular, for each sensor $i$ we train 10 different QRNNs each designed to estimate one specific quantile $a$ from the set [${0.01, 0.1, 0.2, 0.25, 0.5, 0.6, 0.75, 0.8, 0.9, 0.99}$]. More specifically, each $\mathcal{L}_{ai}$ processes data from a single time point $t$ for sensor $i$, denoted as $x_{ti}$, to produce an output $\tilde{q}_a(x_{ti})$. Each $\mathcal{L}_{ai}$ enables precise quantile-based evaluations of the operational status of each sensor at any given moment.




The second step in this block involves training additional QRNNs, denoted as $\mathcal L_{ai}^2$. These networks operate in cascade with the initial $\mathcal L_{ai}$, processing the estimated quantiles \(\tilde{q}_a(x_{ti})\) from the previous step and computing only two new quantiles: \(\tilde{q}_{0.75}(x_{ti})\) and \(\tilde{q}_{0.25}(x_{ti})\).

The second set of QRNNs $\mathcal L_{ai}^2$ is beneficial because it refines the predictions generated by the first set of QRNNs. While the first QRNNs are trained to predict a range of quantiles across the entire distribution of the data, the second set focuses specifically on predicting only the higher or lower extreme quantiles providing more accurate predictions in critical areas of interest, such as extreme values or outliers. 

Each QRNN is a small encoder-decoder network trained using the ADAM optimizer. The architectures for $\mathcal L_{ai}$ and $\mathcal L_{ai}^2$, are summarized below.
The design for $\mathcal L_{ai}$ is:
\begin{itemize}
\item Encoder Layers (No Skip Connections): 4 layers
\item Decoder Layers (No Skip Connections): 4 layers
\item Encoder Nodes: [128, 64, 32, 16] at layers [1, 2, 3, 4]
\item Decoder Nodes: [16, 32, 64, 128] at layers [1, 2, 3, 4]
\end{itemize}
The design for $\mathcal L_{ai}^2$ is:
\begin{itemize}
\item Encoder Layers (With Skip Connections): 2 layers
\item Decoder Layers (With  Skip Connections): 2 layers
\item Encoder Nodes: [32, 16] at layers [1, 2]
\item Decoder Nodes: [16, 32] at layers [1, 2]
\end{itemize}
In our training, we employ Leaky ReLU activation functions and batch normalization across all layers to provide non-linearity and improve input standardization, respectively. We utilize a learning rate of \(1e-3\) and train each model for 300 epochs to balance training time and convergence quality. A dropout rate of 0.1 is chosen to prevent overfitting. For batch normalization, we use an epsilon of \(1e-5\) to ensure numerical stability during training. The momentum parameter for calculating running averages in batch normalization is set to 0.9.

\subsection{Transformer network} 
In the final step of our pipeline, we incorporate a transformer architecture inspired by the framework proposed by \cite{Cholakov}. 
To train the transformer we use the target outcome $y_{ti}$ for each data point $x_{ti}$ of the training set. Each $y_{ti}$ is set to 0 for data points collected under normal operational conditions and 1 for those gathered during abnormal conditions. Additionally, following the approach suggested by \cite{Wu_2021}, we employ the ratio of the 2 quantile outputs from the previous step, as the input for the transformer. This is denoted as \(P(t) = \left| \frac{q_{0.25}}{q_{0.75}} \right|\) and it is adopted because it is a good feature for detecting machine failure (as shown in \cite{Wu_2021}), thereby enhancing the transformer's understanding of the input data.

The architecture of the transformer model is structured with six stacked layers, each incorporating a dual-head self-attention mechanism. This setup enables the model to selectively concentrate on distinct segments of the input sequence, thereby enhancing its ability to capture intricate patterns and dependencies. Each layer includes normalization processes and skip connections. The final task of this transformer is a binary classification where the model differentiates between normal operational data and anomalies that indicate potential system breakdowns.

\begin{table*}[!htb] 
 \centering
 \caption{Ablation Study: F1 Score, Recall, Precision, and Accuracy Obtained by different configurations of our pipeline when predicting machine breakdown over three different time frames (10 Seconds, 30 Minutes, and 1 Hour).}
 \label{tab:ablation}
 \begin{tabular}{c|cccc|cccc|cccc} 

 & \multicolumn{4} {c|} {10 seconds} & \multicolumn{4} {c|}{30 Minutes} &\multicolumn{4} {c}{60 Minutes} \\
 \hline
 Model&F1&Recall&Precision&Acc\%&F1&Recall&Precision&Acc\%&F1&Recall&Precision&Acc\%\\
 \hline
$QR$&86.92&85.47&90.17&85.91&52.71&50.68&54.93&53.82&53.93&52.36&55.61&56.44\\
$QR^2$&99.64&99.60&99.85&99.62&62.40&61.74&63.08&62.11&59.21&58.44&60.01&60.62\\
$Transformer$&99.42&99.24&99.66&99.42&72.82&71.08&74.66&74.34&65.4&64.77&66.06&68.88\\
$All$ &\textbf{99.91}&\textbf{99.95}&\textbf{99.81}&\textbf{99.81}&\textbf{75.90}&\textbf{75.49}&\textbf{76.33}&\textbf{76.63}&\textbf{69.85}&\textbf{69.18}&\textbf{70.55}&\textbf{70.84}\\
 \end{tabular}
\end{table*}

\begin{table*}[!htb] 
 \centering
 \caption{Comparison Study: F1 Score, Recall, Precision, and Accuracy computed for the proposed method against other approaches when predicting machine breakdown over three different time frames (10 Seconds, 30 Minutes, and 1 Hour).}
 \label{tab:comparison}
 \begin{tabular}{c|cccc|cccc|cccc} 

 & \multicolumn{4} {c|} {10 seconds} & \multicolumn{4} {c|}{30 Minutes} &\multicolumn{4} {c}{60 Minutes} \\
 \hline
 Model&F1&Recall&Precision&Acc\%&F1&Recall&Precision&Acc\%&F1&Recall&Precision&Acc\%\\
 \hline
SVR&95.11&96.08&96.74&95.85&67.02&66.09&68.03&68.88&62.04&59.83&61.89&62.33\\
KNN&99.32&99.21&98.81&99.58&61.78&60.95&63.14&61.88&59.67&58.36&60.58&60.71\\
LSRN&98.44&98.67&99.08&99.38&73.33&71.19&73.82&73.49&65.11&64.83&67.08&68.17\\
LS-SVM&99.18&99.54&99.36&99.3&69.71&67.82&69.94&70.29&66.14&65.37&66.67&68.04\\
\cite{Ma}&87.09&84.67&89.92&85.39&53.69&50.16&55.28&52.93&54.66&52.18&55.72&56.91\\
\cite{Grigsby}&99.42&99.24&99.66&99.42&72.82&71.08&74.66&74.34&65.40&64.77&66.06&68.88\\
Proposed&\textbf{99.91}&\textbf{99.95}&\textbf{99.83}&\textbf{99.82}&\textbf{75.90}&\textbf{75.49}&\textbf{76.33}&\textbf{76.63}&\textbf{69.85}&\textbf{69.18}&\textbf{70.55}&\textbf{70.84}\\
 \end{tabular}
\end{table*}

\setlength{\tabcolsep}{3pt}

\begin{figure*}[t]
\begin{center}
\includegraphics[width=0.9\textwidth,trim={0cm 3cm 2cm 0cm},clip]{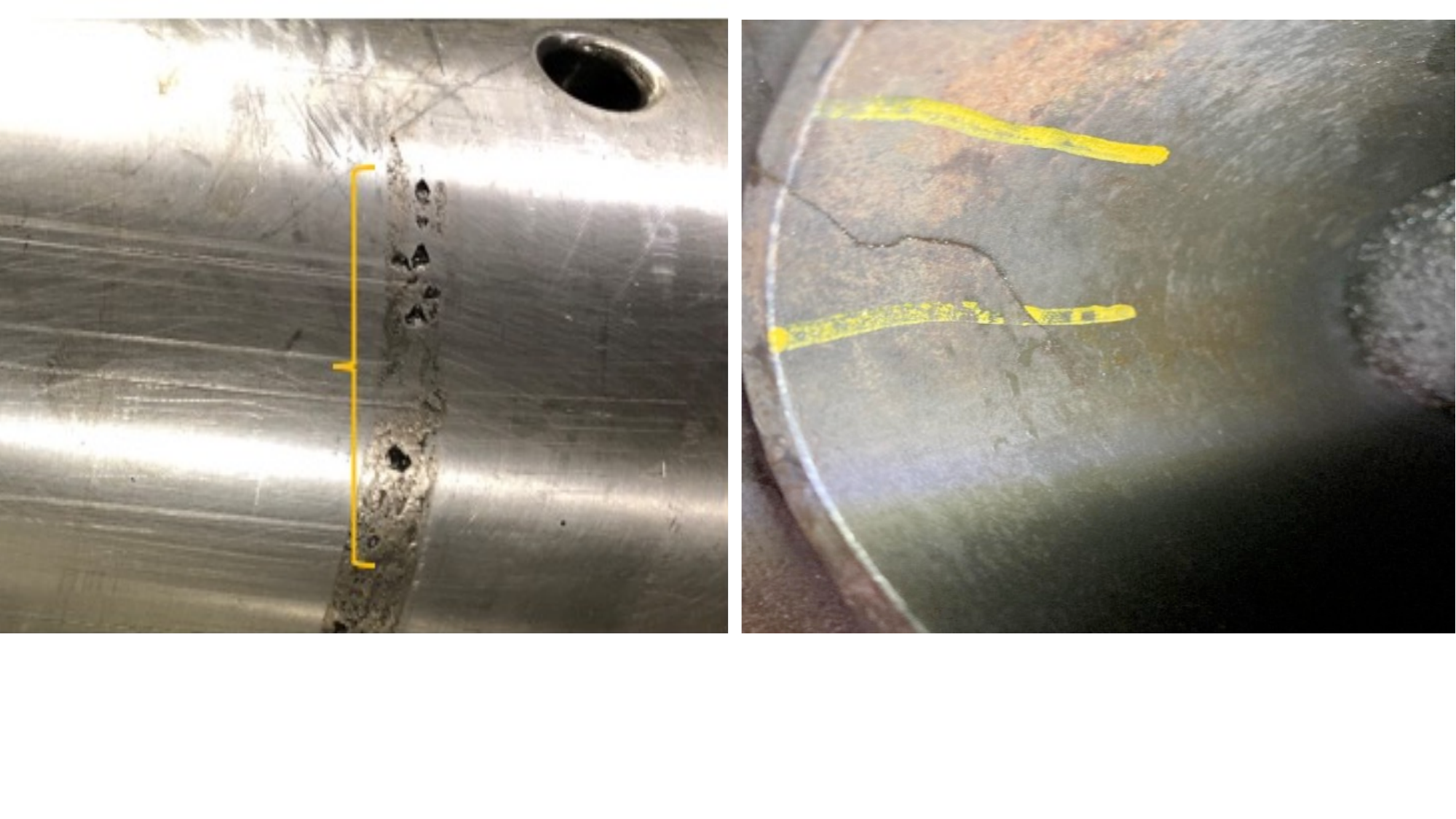}
\caption{Examples of damaged mechanical components, such as punch sleeves and swing levers, were predicted by our proposed approach before the damage occurred on two separate manufacturing machines.\label{fig:damaged}}
\end{center}
\end{figure*}

\section{Experiment Results}\label{experiment}
We have evaluated our proposed pipeline within a real-world manufacturing environment, demonstrating its capability to predict system breakdown. In particular, in Section~\ref{sec:metrics} we present the metrics used in our evaluation, in Section~\ref{sec:ablation} we conduct an ablation study to evaluate the contribution of each component within our pipeline. In Section~\ref{sec:comparison}, we present a comparative analysis of various machine learning applications against our proposed method. In Section~\ref{sec:computational_cost}, we assess the computational time performance to confirm the real-time capabilities of our approach. Finally, Section~\ref{sec:downtown} outlines an estimation of the savings that would be achieved through the implementation of our research.

\subsection{Evaluation Results and Metrics}\label{sec:metrics}
To evaluate our pipeline, we have chosen to utilize a set of quantitative metrics that include precision, recall, F1 score, and accuracy. These metrics are calculated over different lengths of time sequences: 10 seconds, 30 minutes, and 1 hour. The rationale behind this selection is to assess the model's performance across a spectrum of short to long-term predictions. Making predictions over longer time frames is inherently more challenging due to the increased complexity involved. By evaluating the model across these distinct time intervals, we aim to understand its predictive capabilities and reliability in real-world operational scenarios. All these metrics allow us to gauge not only the model's overall accuracy but also understand its precision in identifying true positives, its recall rate in capturing all relevant instances, and the balance between precision and recall as indicated by the F1 score. Below, we have summarized the formulations for the considered metrics.
\begin{equation}
   \text{Precision}=\frac{\text{True Positives}}{\text{True Positives}+\text{False Positives}}
\end{equation}

\begin{equation}
\text{Recall}=\frac{\text{True Positives}}{\text{True Positives}+\text{False Negatives}}
\end{equation}

\begin{equation}
\text{F1}=\frac{2*\text{Precision}*\text{Recall}} {\text{Precision}+\text{Recall}}
\end{equation}

\subsection{Ablation study}\label{sec:ablation}
The results of our ablation study are presented in Table \ref{tab:ablation}, which analyzes the effective contribution of each component of our pipeline. The configurations considered in our study are: QR, which uses only the first set of QRNNs $\mathcal L_{ai}$; \(QR^2\), which uses both $\mathcal L_{ai}$ and $\mathcal L_{ai}^2$; Transformer, that proposes passing the raw data directly to the transformer network and All which include the entire pipeline.

The QR configuration achieved the lowest accuracy of 85\% for the shorter 10-second prediction timeframe, and its performance drastically reduced to 53\% and 56\% accuracy when considering longer timeframes of 30 minutes and 60 minutes, respectively.

The \(QR^2\) configuration performed slightly better, with a good accuracy of 99\% for a short period of 10 seconds but still dropping performance to 62\% and 60\% for longer predictions. The transformer network seems to provide the largest contribution, achieving 99\% accuracy on the short period and experiencing a reduced drop in performance to 74\% and 68\%. The configuration that uses all components (last row in Table \ref{tab:ablation}) provided the best performance, delivering the highest level of performance with 99\% accuracy on the short period and maintaining a relatively high accuracy of 76\% and 70.84\% for longer predictions of 30 minutes and 60 minutes, respectively.

\subsection{Comparison against related state-of-the-art}\label{sec:comparison}
In Table \ref{tab:comparison}, we compared our pipeline, TQRNNs, against a range of state-of-the-art models. These included standard machine learning approaches such as i) Support Vector Regression (SVR), ii) K-Nearest Neighbor (KNN), iii) Least Square Residual Network (LSRN), and iv) Least Square Support Vector Machine (LS-SVM), as well as more advanced techniques cited as \cite{Grigsby} and \cite{Ma}. Also for this experiment, we conducted evaluations across three different intervals: 10 seconds, 30 minutes, and 1 hour. All models were tested under the same conditions and within the same live manufacturing environment.

When predicting anomalies within a short period of 10 seconds, all models performed relatively well, achieving an accuracy greater than 90\%. However, the performance of state-of-the-art approaches significantly declined when the prediction time frame was increased. Our approach mitigated this problem and achieved the highest accuracy for both the 30-minute and 60-minute intervals, with 76\% and 70\% accuracy, respectively. This confirms its robustness and underscores the advantages of our solution compared to other methods.

From the table, we can also note the relatively good performance of our pipeline in comparison with the approaches proposed by \cite{Grigsby}. For the 1-hour interval, \cite{Grigsby} achieved a lower accuracy of 68.88\%, while our method exceeded it by a margin of 1.96\%. Moreover, when compared against the framework proposed by \cite{Ma}, which achieved an accuracy of 56.91\%, our pipeline exhibited a substantial margin of improvement, outperforming it by 13.93\%.

Finally, in Figure \ref{fig:damaged}, we show an example of damage in two critical machine components identified by our pipeline 60 minutes before the actual damage occurred. The early detection capability of our system, is crucial, as it enables preemptive maintenance. This eliminates the need to halt production for damaged part replacements, which would have also led to the expensive waste of producing faulty products.

\subsection{Computational cost analysis}\label{sec:computational_cost} 
In our experiment, we utilized a workstation equipped with an AMD EPYC 7763 processor, featuring 64 cores, to measure the computational times for each block of the pipeline. The obtained computational times are as follows: 413.2 milliseconds for the QRNNs $\mathcal{L}{ai}$, 475.4 milliseconds for the QRNNs $\mathcal{L}{ai}^2$, and 586.16 milliseconds for the final transform block. The total time taken by the processing pipeline is 1474.76 milliseconds, which means we can run the processing in real-time, approximately once per second. In our implementation, we make use of the   PyTorch library for neural network operations and sequence processing. These tools were integral in optimizing the performance and efficiency of our processing pipeline, particularly in handling the high computational demands and ensuring the real-time capability of our system. This selection facilitated efficient GPU utilization and optimal threading across the processor cores, essential for experimental reproducibility and scalability.

\subsection{Machine downtime}\label{sec:downtown}
In our experiment, we also assessed the reduction in machine downtime achieved by integrating our solution into the production process. We discovered that our solution decreased downtime from 60 hours per month to approximately 5 hours. This significant enhancement in system reliability led to an increase in operational efficiency from 78.38\% to 89.62\%. As a result, this boost in efficiency translates to an additional monthly production of 1,689,600 cans. These findings highlight the substantial impact our model can have on enhancing production efficiency and boosting output.

\section{Conclusion}\label{conclusion}
In this study, we introduce TQRNNs a novel solution for real-time machine failure prediction in manufacturing. TQRNNs consist of a set of QRNNs designed to efficiently segment data, along with a concatenated transformer to ensure accurate classification over extended timeframes of up to one hour. We evaluated our model within the beverage cans industry and found that it maintains high accuracy and enhances outlier prediction compared to state-of-the-art methods.

Specifically, TQRNNs enable timely machine stoppages and prevent the production of defective products by predicting equipment failures up to one hour in advance. 

We believe that our model has broad applicability across various industrial sectors. When deployed on a cloud system, it can provide a flexible and scalable solution. It can be used to detect anomalies, reduce machine downtime, boost productivity, improve equipment effectiveness, ensure continuous production, and finally lower costs by minimizing resource waste and financial losses from defective products. Future enhancements to this pipeline will focus on extending predictions up to 48 hours ahead in real-time. Additionally, we plan to broaden our research to encompass multiple machines and applications, demonstrating the robustness and versatility of our development.

\section*{Acknowledgment}
This research was conducted within a beverage cans manufacturing facility. Lemuel Puglisi is enrolled in a PhD program at the University of Catania, fully funded by PNRR (DM 118/2023).


\begin{thebibliography}{00}

\bibitem{Frank} Frank, Alejandro Germán, Lucas Santos Dalenogare, and Néstor Fabián Ayala. "Industry 4.0 technologies: Implementation patterns in manufacturing companies." International journal of production economics 210 (2019): 15-26

\bibitem{Monia} Monia Niero, Stig Irving Olsen, "Circular economy: To be or not to be in a closed product loop? A Life Cycle Assessment of aluminium cans with inclusion of alloying elements," Resources, Conservation and Recycling, Volume 114, 2016

\bibitem{Li} Li, Yao, Yihai He, Ruoyu Liao, Xin Zheng, and Wei Dai. "Integrated predictive maintenance approach for multistate manufacturing system considering geometric and non-geometric defects of products." Reliability Engineering and System Safety 228 (2022)

\bibitem{Lai} Lai, Chin-Feng, Wei-Che Chien, Laurence T. Yang, and Weizhong Qiang. "LSTM and edge computing for big data feature recognition of industrial electrical equipment." IEEE Transactions on Industrial Informatics 15, no. 4 (2019): 2469-2477

\bibitem{Serradilla} Serradilla, Oscar, Ekhi Zugasti, Jon Rodriguez, and Urko Zurutuza. "Deep learning models for predictive maintenance: a survey, comparison, challenges and prospects." Applied Intelligence 52, no. 10 (2022): 10934-10964.

\bibitem{Ma} Ma, Meng, and Zhu Mao. "Deep-convolution-based LSTM network for remaining useful life prediction." IEEE Transactions on Industrial Informatics 17, no. 3 (2020): 1658-1667

\bibitem{Yurek} Yurek, Ozlem Ece, and Derya Birant. "Remaining useful life estimation for predictive maintenance using feature engineering." In 2019 Innovations in Intelligent Systems and Applications Conference (ASYU), pp. 1-5. IEEE, 2019

\bibitem{Yang} Yang, Tao, Jinming Wang, Weijie Hao, Qiang Yang, and Wenhai Wang. "Hybrid Cloud-Edge Collaborative Data Anomaly Detection in Industrial Sensor Networks." arXiv preprint arXiv:2204.09942 (2022)

\bibitem{Grigsby} Grigsby, Jake, Zhe Wang, Nam Nguyen, and Yanjun Qi. "Long-range transformers for dynamic spatiotemporal forecasting." arXiv preprint arXiv:2109.12218 (2021)

\bibitem{Huynh} Huynh, Khac Tuan, Antoine Grall, and Christophe Bérenguer. "A parametric predictive maintenance decision-making framework considering improved system health prognosis precision." IEEE Transactions on Reliability 68, no. 1 (2018): 375-396

\bibitem{Hanshuo} Hanshuo, Mu, Zhai Xiaodong, Tu Xuan, and Qiao Fei. "Research on fault prediction method of electronic equipment based on improved SVR algorithm." In 2020 Chinese Automation Congress (CAC), pp. 3092-3096. IEEE, 2020

\bibitem{Bakir} Bakir, A. A., M. Zaman, A. Hassan, and M. F. A. Hamid. "Prediction of remaining useful life for mechanical equipment using regression." In Journal of Physics: Conference Series, vol. 1150, no. 1, p. 012012. IOP Publishing, 2019

\bibitem{Xiao} Xiao, Yi, Hongru Ren, Renquan Lu, and Shen Cheng. "Manufacturing Big Data Modeling Based on KNN-LR Algorithm and Its Application in Product Design Business Domain." In 2021 IEEE 10th Data Driven Control and Learning Systems Conference (DDCLS), pp. 367-371. IEEE, 2021

\bibitem{Rivas} Rivas, Alberto, Jesús M. Fraile, Pablo Chamoso, Alfonso González Briones, Inés Sittón, and Juan M. Corchado. "A predictive maintenance model using recurrent neural networks." In 14th International Conference on Soft Computing Models in Industrial and Environmental Applications (SOCO 2019) Seville, Spain, May 13–15, 2019, Proceedings 14, pp. 261-270. Springer International Publishing, 2020

\bibitem{Chen_2021} Chen, Chuang, Zheng Hong Zhu, Jiantao Shi, Ningyun Lu, and Bin Jiang. "Dynamic predictive maintenance scheduling using deep learning ensemble for system health prognostics." IEEE Sensors Journal 21, no. 23 (2021): 26878-26891

\bibitem{Devlin} Devlin, Jacob, Ming-Wei Chang, Kenton Lee, and Kristina Toutanova. "Bert: Pre-training of deep bidirectional transformers for language understanding." arXiv preprint arXiv:1810.04805 (2018)

\bibitem{Batur} Batur, Demet, Jennifer M. Bekki, and Xi Chen. "Quantile regression metamodeling: Toward improved responsiveness in the high-tech electronics manufacturing industry." European Journal of Operational Research 264, no. 1 (2018): 212-224

\bibitem{Korkmaz} Korkmaz, Mustafa Ç., and Christophe Chesneau. "On the unit Burr XII distribution with the quantile regression modeling and applications." Computational and Applied Mathematics 40, no. 1 (2021): 29

\bibitem{Tan} Tan, Kian Long, Chin Poo Lee, Kalaiarasi Sonai Muthu Anbananthen, and Kian Ming Lim. "RoBERTa-LSTM: a hybrid model for sentiment analysis with transformer and recurrent neural network." IEEE 

\bibitem{Cannon} Cannon, Alex J. "Quantile regression neural networks: Implementation in R and application to precipitation downscaling." Computers and geosciences 37, no. 9 (2011): 1277-1284

\bibitem{Biau} Biau, Gérard, and Benoît Patra. "Sequential quantile prediction of time series." IEEE Transactions on Information Theory 57, no. 3 (2011): 1664-1674

\bibitem{Mohammad} Mohammad Sina Jahangir, John You, John Quilty, "A quantile-based encoder-decoder framework for multi-step ahead runoff forecasting," Journal of Hydrology, Volume 619, 2023


\bibitem{Cholakov} Cholakov, Radostin, and Todor Kolev. "Transformers predicting the future. Applying attention in next-frame and time series forecasting." arXiv preprint arXiv:2108.08224 (2021)

\bibitem{Wu_2021} Wu, Haixu, Jiehui Xu, Jianmin Wang, and Mingsheng Long. "Autoformer: Decomposition transformers with autocorrelation for long-term series forecasting." Advances in Neural Information Processing Systems 34 (2021): 22419-22430

\end{thebibliography}
\end{document}